\documentstyle[prb,preprint,aps]{revtex}
\tightenlines 
\begin{document}
\draft

\title{\bf Statistics of Raman-Active Excitations Via \\
Measurement of Stokes--Anti-Stokes Correlations}

\author{\"{O}zg\"{u}r E. M\"{u}stecapl{\i}o\u{g}lu and Alexander S.
Shumovsky}

\address{Physics Department, Bilkent University, Bilkent, 06533 Ankara, Turkey}

\maketitle

\begin{abstract}
A general fundamental relation connecting the correlation
of Stokes and anti-Stokes modes to the quantum statistical
behavior of vibration and pump modes in Raman-active
materials is derived. We
show that under certain conditions this relation can be
used to determine the equilibrium number variance of phonons.
Time and temperature
ranges for which such conditions can be satisfied are studied
and found to be
available in todays' experimental standards. Furthermore, we  
examine the results in the presence of multi-mode pump as well as 
for the coupling of pump
to the many vibration modes and discuss their validity in these
cases.  

\end{abstract}
\narrowtext
\newpage
\section{INTRODUCTION}
The concept of squeezed state has been established in the 
language of physics mainly by the developments in quantum optics. 
On the other hand, basic requirement of finding a system in a squeezed 
state is to have bosons as the constituents of the system interacting in 
a pairwise manner and that might be fulfilled not only in optical systems 
but in some other Bose-type systems as well. In actual fact, the 
introduction of squeezed states in optics \cite{stol} was based on the 
previous consideration of superfluidity\cite{bog} in liquid $He^4$ 
(also see \cite{rob}). While squeezing of quantum fluctuations is 
the most well-known aspect of squeezed states, rich variety of effects 
might be expected due to their interesting statistical properties even 
at thermal equilibrium. Certain effects like anti-bunching have already 
been observed in the realm of quantum optics and this makes it an
intriguing question how to find squeezed states and their effects in other
places. In this context, few proposals have been suggested for the generation
and detection of squeezed states of Bose-type excitations in 
solids\cite{art,hu,mus}. Quite recently, squeezed phonons have been 
produced and detected\cite{gar}.

It is very interesting that, unlike the case of light, the squeezed states
of phonons may arise from different microscopic interactions in solids
even at thermal equilibrium \cite{shu98}. Deviations from typical 
equilibrium distribution of phonons, namely Bose-Einstein distribution,
might  arise from anharmonic interactions among phonons or
from some other mechanisms such as the polariton coupling in
ionic crystals \cite{art,shu91} or polaron
mechanism \cite{alt}. In such cases, equilibrium distribution
of phonons are  that of squeezed thermal phonons\cite{kim}. 
Therefore, it seems to be an important question how to determine
the equilibrium distribution of phonons when there is a possibility
that phonons can be found to be in non-classical states. As a particular 
example of some considerable interest, the squeezed states of phonons due 
to the photon - optical phonon interaction in an ionic crystal \cite{shu91} 
should be mentioned here. The polariton coupling in such a system is 
described by the following Hamiltonian \cite{mad}
\begin{eqnarray}
H & = & \frac{1}{2} \sum_k H_k, \nonumber \\
H_k & = & \omega_k a^{\dagger}_ka_k+ \omega_b b^{\dagger}_kb_k +
ig_k[(a^{\dagger}_k-a_{-k})(b^{\dagger}_k+
b_{-k})+(a^+_{-k}-a_k)(b_{-k}+b^+_k)] \nonumber
\end{eqnarray}
where $\omega_k$ is the photon frequency, $\omega_b$ is the frequency of 
transversal oscillations of optical phonons, $g_k$ is the polariton 
coupling constant and the operators $a_k,b_k$ describe the annihilation of 
photons and optical phonons respectively. Since the Hamiltonian under 
consideration is the Hermitian bilinear form, it can be diagonalized by 
the Bogolubov canonical transformation \cite{bog} similar to that used in 
the definition of squeezed states \cite{stol}. As a result, the thermal 
equilibrium state of the system is described by the following density 
matrix
\begin{eqnarray}
\rho( \beta )= \frac{e^{- \beta H_p}}{Tr e^{- \beta H_p}} \nonumber
\end{eqnarray}
where $H_p$ denotes the Hamiltonian $H$ in diagonal (polariton) 
representation and $\beta$ is the reciprocal temperature. In analogy to 
the quantum optics, consider the so-called degree of coherence \cite{man}
\begin{eqnarray}
G^{(2)}= \frac{\langle b^{\dagger 2}b^2 \rangle}{\langle b^{\dagger}
b \rangle} \nonumber
\end{eqnarray}
where $\langle ... \rangle$ denotes the average with respect to the density 
matrix $\rho ( \beta )$. It is straightforward to calculate $G^{(2)}$ as a 
function of temperature for typical parameters of an ionic crystal (see Fig. 
1). One can see that, at low temperatures, $G^{(2)} \approx 8$, while the 
same correlation function calculated with the Bose-Einstein distribution 
gives $G^{(2)}_{BE}=2$. It is also seen that the strong quantum fluctuations 
can be observed only below $T\sim 50K$ because they are eroded by thermal 
fluctuations with the increase of temperature.   

In contrast to the case of non-classical states of photons  
there is no an efficient direct method of measurement 
allowing the characterization of the
quantum state of Bose-type excitations in solids\cite{hu}. 
Even though correlation functions to any
order would be demanded to describe fully a quantum state, it is
usually good enough to distinguish quantum states by their number
variances\cite{man}. 
Here, we present a way
to determine the number variance of phonons at equilibrium 
in a Raman active medium. It is already suggested that
correlation Raman spectroscopy may be used to measure the quantum
statistical properties of a vibration mode for the case of 
Stokes (S) type Raman scattering through a measurement of 
the intensity and the Mandel's
Q-factor of the Rayleigh mode\cite{shu93}. However, even at low 
temperatures vacuum fluctuations of the Anti-Stokes (AS) modes 
might disturb measurements of high order correlations and thus 
careful study of the role of the AS modes in such measurements is 
demanded. In this article, we follow
a similar ideology in more general terms by examining both the S and 
AS components of multi-mode Raman scattering. Even though the problem 
becomes analytically intractable when AS modes are included, it is now 
possible to establish an interesting connection between the number 
variance of phonons and the correlations of S and AS modes. Moreover, due
to the removing low temperature restriction in the exclusion of AS
modes, influence of temperature in the high order quantum correlations 
can be examined as well.

The paper is outlined as follows. In Sec. II, using a general model
of Raman type three-body scattering, we find the inter-mode correlation
function of S and AS modes. Discussion of this general result
under standard approximations of Raman scattering, with an
emphasis of modifications in their range of validity, is subject to
Sec.III. Finally, Sec.IV gives a brief summary of our results and 
conclusions.
\section{Correlation of Stokes and anti-Stokes photons}
General relations between
the correlation function of S and AS modes and the number variance of
phonons is developed in this section for the following Raman-type
Hamiltonian,
\begin{eqnarray}
H=\sum_{{\bf k}\lambda}\omega_{{\bf
k}\lambda}a^{\dagger}_{{\bf k}\lambda} a_{{\bf k}\lambda}+\sum_{{\bf
kk^{\prime}q}}(M^{S}_{{\bf kk^{\prime}q}} a^{\dagger}_{{\bf
k^{\prime}}S}a_{{\bf k}R}a^{\dagger}_{{\bf q}V}+ M^{A}_{{\bf
kk^{\prime}q}}a^{\dagger}_{{\bf k^{\prime}}A} a_{{\bf k}R}a_{{\bf
q}V}+H.c.),
\end{eqnarray}
where $a^{\dagger}_{{\bf k}\lambda}(a_{{\bf k}\lambda})$ are the
creation (annihilation) operators for the $\lambda$-mode with
momentum ${\bf k}$ and
corresponding frequency $\omega_{{\bf k}\lambda}$. Here the mode
index $\lambda=S,A,V,R$ stands for Stokes, Anti-Stokes, vibration and
Rayleigh modes, respectively. As usually, the polarization labels are
suppressed within the momentum symbols for the sake of notational
simplicity.  Coupling constants are denoted by $M^{S}_{{\bf
kk^{\prime}q}}$ for the S-type scattering and $M^{A}_{{\bf
kk^{\prime}q}}$ for the AS-type scattering.
While writing this tri-linear bosonic Hamiltonian we assumed as 
usually \cite{she} that the Raman scattering is observed under the 
condition $\omega_{R,S,A} \gg \omega_V$ when the pair-wise creation
of radiation modes has quite small probability so that energy is conserved. 
This supposition is equivalent to the rotating wave approximation of
the quantum optics\cite{all}.
We also assumed that the radiation consists of three $R$, $S$, and 
$AS$ pulses which are well-separated
on the frequency domain so that $[a_{{\bf
k}\lambda},a^{\dagger}_{{\bf k^{\prime}}\lambda^{\prime}}]
=\delta_{{\bf kk^{\prime}}}\delta_{\lambda\lambda^{\prime}}$.
If a single-mode strong coherent (classical) pumping is assumed, 
all one can expect is that the phase-matching conditions would have 
limited the number of active phonon modes to one. Nevertheless, it seems
to be reasonable to consider the Raman scattering by an infinite
Markoffian system of phonons\cite{wal1,per1}. In particular, it permits 
oneself to take into account the broadening of $S$ and $AS$ lines.   
The usual selection rules of Raman scattering, 
namely phase-matching or quasi-resonance conditions\cite{she}, are not 
essential for the derivation of the general relations below. 
Therefore, the results given in this section are also valid in not so 
perfect Raman coupling situations which should be important in real 
materials.

If we define the number operator $n_{{\bf k} \lambda}$ for the 
$\lambda$-mode with momentum ${\bf k}$ as $n_{{\bf k} \lambda}=a_{{\bf
k}\lambda}^{\dagger}a_{{\bf k}\lambda}$, then the total number
operator $N_{\lambda}$ for $\lambda$-mode becomes $N_{\lambda}
=\sum_{{\bf k}}n_{{\bf k}\lambda}$. Heisenberg equations of motion
yield the conservation laws, also known as Manley-Rowe
relations\cite{she},
\begin{eqnarray}
N_{S}+N_{A}+N_{R}=C_{1}, \\\nonumber
N_{S}-N_{A}-N_{V}=C_{2}.
\end{eqnarray}
Here constant operators $C_{1}, C_{2}$ are specified by the initial
conditions. Similar relations can also be constructed for the scattering 
of photons of a monochromatic laser beam from a dispersionless optical
phonon\cite{car,shu98}.
Solving these equations for $N_{S}$ and $N_{A}$, the S and AS
correlation function is found to be
\begin{eqnarray}
<N_{A};N_{S}>=\frac{1}{4}(V(C_{1})-V(C_{2})+V(N_{R})-V(N_{V})
-2<C_{1};N_{R}>-2<C_{2};N_{V}>),
\end{eqnarray}
where the correlation function $<A;B>$ of two operators $A,B$
is defined by
\begin{eqnarray}
<A;B>=<AB>-<A><B> \nonumber
\end{eqnarray}
and hence  variance of operator $A$ is
given by the self-correlation function $V(A)=<A;A>$. Here the
averages $<.>$ are with respect to the initial
state since Heisenberg picture is used. It is natural to consider
an initial state in which the $S$ and $AS$ modes are in their vacuum
states, when we obtain,
\begin{eqnarray}
<N_{A}(t);N_{S}(t)>&=&\frac{1}{4}(V(N_{R}(0))-V(N_{V}(0))+V(N_{R}(t))-
V(N_{V}(t)) \nonumber \\
&-&2<N_{R}(0);N_{R}(t)>-2<N_{V}(0);N_{V}(t)>).
\end{eqnarray}
An operator $A$ at time $t$ is indicated by $A(t)$ while initially by 
$A(0)$. That equation connects the S and AS correlation function to the
quantum statistical behavior of phonons and pump photons.

Within conventional Raman theory quantum properties of pump are
usually neglected through the classical pump
assumption\cite{wal2,lou}.  This approximation introduces a time range
to the problem during which changes in the pump intensity remains
negligible.  We can apply a similar approximation by assuming an
intense laser pump with photons in coherent states and performing a
mean field average over them in the above equations.  Under this
assumption, the correlation function of the S and AS modes is related
only to phonon statistics and the initial, known, number variance of
the pump photons. However, time range of validity for the parametric
approximation should be modified in our case. As we shall show in the
subsequent section, statistical behavior of the pump might change
significantly in shorter time than the occurance of a significant
change in its intensity.  Our purpose is to examine the equilibrium
statistics of phonons determined by $V(N_{V}(0))$; therefore we need
to express all time dependent terms on the right hand side of the
Eq.4 in terms of initial operators to see any further relation
between the S and AS correlation function and the equilibrium
variance of phonons. For that aim we specify a model system and study
its dynamics. 

We conclude this section by noting that a similar
relation can be derived for the molecular Raman model, which is
equivalent to the full bosonic Raman model under Holstein-Primakoff
approximation in the case of low excitation density\cite{cho}. In
that case, S and AS correlations depend on the quantum statistics of
population distributions of the molecular energy levels.

\section{DISCUSSIONS FOR PARAMETRIC RAMAN MODEL}

In reality, coupling
of one vibration mode to the pump beam for sufficiently long time
of measurement is not an easy task.  Therefore, in this section we
investigate a Raman scattering in which coupling of pump photons to
all phonon modes are allowed.  We shall treat the pump as an intense
coherent beam of photons and thus its state $\mid\psi_{R}\rangle$
in general is described by a multimode coherent state,
\begin{eqnarray}
\mid\psi_{R}\rangle=\prod_{\bf l}\otimes\mid\alpha_{\bf l}\rangle
\end{eqnarray}
in which $\alpha_{\bf l}$ are the coherence parameters of the 
modes ${\bf l}$.
According to the remarks at the end of previous section, we now 
perform mean field averaging with respect to pump photon states
in Eq.1 assuming the Raman-active material is placed in an
ideal cavity which selects single modes for S and AS radiations,
namely ${\bf k^{\prime} = k_{A,S}}$.
Then after dropping constant terms the Hamiltonian in Eq.1 reduces to
an effective one, \begin{eqnarray}
H^{eff}=\sum_{\lambda=S,A}\omega_{\lambda}n_{\lambda}+
\sum_{{\bf q}}\omega_{{\bf q}V}a^{\dagger}_{{\bf q}V}a_{{\bf q}V}+
\sum_{{\bf q}}(
g_{{\bf q}}^{S}a^{\dagger}_{S}a^{\dagger}_{{\bf q}V}+g_{{\bf q}}^{A}
a^{\dagger}_{A}a_{{\bf q}V}+H.c.),
\end{eqnarray}
where new effective coupling constants $g_{{\bf q}}^{A,S}$ are 
introduced by
\begin{eqnarray}
g_{{\bf q}}^{A,S}=\sum_{\bf k}M_{{\bf kk_{A,S}q}}^{A,S}
\alpha_{{\bf k}}
\end{eqnarray}
The summation above can be calculated once the density of states
for the pump is also specified. As one can see, the Hamiltonian
will be in the given form, involving summations over phonon modes, 
in all cases except the case of perfectly phase matched single pump 
and phonon modes. In order to make sure that our results are not
too susceptible to any imperfectness of the system arising from
multi-mode nature of pump or phase-mismatches among the phonon
and photon modes, we shall treat the problem using the model 
described by the above Hamiltonian involving summations over phonon 
modes. 
When finite number of phonon modes are assumed, which is reasonable
for real crystals of finite size, then
such a model becomes integrable since the dynamics is ruled by the
following closed set of operator linear differential equations,
\begin{eqnarray}
i \frac{d}{dt}a_{{\bf q}V} &=& \omega_{{\bf q}V}a_{{\bf q}V} +
g^{S}_{\bf q}a^{\dagger}_{S} + g^{A\ast}_{{\bf q}}a_{A},
                     \nonumber \\
i\frac{d}{dt}a_{S}^{\dagger} &=&  -\omega_{S}a_{S}^{\dagger}-
\sum_{{\bf q}}g^{S\ast}_{{\bf q}}a_{{\bf q}V}, \\
i\frac{d}{dt}a_{A} &=& \omega_{A}a_{A}+\sum_{{\bf q}}g^{A}_{{\bf q}}
a_{{\bf q}V}. \nonumber
\end{eqnarray}
Let us introduce a vector of operators such that
$Y=[a_{S}^{\dagger},a_{A},\{a_{{\bf q}V}\}]^{T}$. We denote
the matrix of coefficients in the above set of equations by $M$ and
its diagonalizing matrix by $D$, so that $D^{-1}MD=E{\bf 1}$ with
eigenvalues $E$. Thus, we get
\begin{eqnarray}
Y_{i}(t)=D_{ij}D^{-1}_{jk}Y_{k}(0)\exp{(-iE_{j}t)},
\end{eqnarray}
where summation over repeated index is implied. It is therefore
possible to write the solution for $\lambda=S,A$-modes in the form,
\begin{eqnarray}
a_{\lambda}(t)^{\dagger}=u_{\lambda}(t)a_{S}^{\dagger}+v_{\lambda}(t)
a_{A}+\sum_{\bf q}w_{{\bf q}\lambda}(t)a_{{\bf q}V}.
\end{eqnarray}
Operators without time arguments are taken at $t=0$. Time dependent
parameters $u,v,w$ are determined by the elements of matrix $D$ and
eigenvalues $E$. Let us
note here that some general relations exists among $u,v,w$ due to the
commutation relations for $a_{\lambda}$ operators and they are not
independent each other.  More explicit way of evaluating $u,v,w$ is
presented below for the single mode phonon case where vector $Y$
reduces to three dimensions in operator space.  When there are no
scattered light modes initially, the correlation function of $S$ and
$AS$ modes becomes
\begin{eqnarray}
\langle n_{S}(t);n_{A}(t)\rangle=A(t)+\sum_{{\bf kq}}B_{{\bf kq}}(t)
\langle a_{{\bf k}V}^{\dagger}a_{{\bf q}V}\rangle+
\sum_{{\bf klpq}}C_{{\bf klpq}}(t)\langle a^{\dagger}_{{\bf k}V}
a_{{\bf q}V}; a^{\dagger}_{{\bf l}V}a_{{\bf p}V}\rangle.
\end{eqnarray}
Here, parameters $A,B,C$ are functions of $u,v,w$.
Since the summations above can be converted into integrals involving
phonon density of states, we see that if there are Van Hove singularities
corresponding to the modes selected by Raman scattering, as in the case of
recent experiments on the generation of non-classical phonon states via
Raman scatterings\cite{gar}, then the
correlation of $S$ and $AS$ modes will be determined strongly by that
mode. If this not the case, then one can still expect domination of the
modes obeying Raman selection rules. Then for that mode the random
phase approximation permits us to write\cite{hau}
\begin{eqnarray}
\langle n_{S}(t)\rangle&=&\mid v_{S}(t)\mid^{2}+
\mid w_{S}^{\prime}\mid^{2}(1+n_{V})\\\nonumber
\langle n_{A}(t)\rangle&=&\mid u_{A}(t)\mid^{2}+
\mid w_{A}^{\prime}\mid^{2}n_{V}\\\nonumber
\langle n_{S}(t);n_{A}(t)\rangle&=&A^{\prime}(t)+B^{\prime}(t)
n_{V}+C^{\prime}(t)V(n_{V}),
\end{eqnarray}
in which the momentum label corresponding to relevant mode is fixed and
dropped for the notational simplicity and primed parameters evaluated at
that mode. It is possible to argue by the results above that a measurement
of the correlation between S and AS can be utilized to determine the
variance of vibration modes, which we usually consider as phonons here,
provided one knows the mean number of such modes initially. The latter
information can be determined by either one of the first two relations
in Eq. 12 after measurement of radiation
mode intensities. Also measurement of radiation mode intensities and
the knowledge of initial phonon number allow
one to keep track of the evolution of mean phonon number through 
the Manley-Rowe relations given by the Eq.(2). Interestingly, since
the mean number of phonons with non-classical distributions deviate
significantly from that of Bose-Einstein distribution, it might
be possible to find some traces of non-classicality even here.
However, in order to classify the distribution of phonons strictly
it would still be necessary to find the next moment of the distribution, 
in other words the variance of phonons. 

Now, an explicit way of determining $u,v,w$ parameters will be
demonstrated for the case of a single phonon mode.
Because of three dimensional operator space in this situation,
eigenvalues $E_{l}$ are found to be as the roots of the cubic equation
\begin{eqnarray}
E^{3} &+& 3\omega_{V}E^2-[\omega_{R}^{2}-3\omega_{V}^{2}+(|g^{A}|^2-
|g^{S}|^{2})]E+
[|g^{S}|^{2}(\omega_{R}+\omega_{V})+|g^{A}|^{2}(\omega_{R}-\omega_{V})]+
\nonumber\\
&+& \omega_{V}(\omega_{V}^{2}-\omega_{R}^{2})=0. \nonumber
\end{eqnarray}
Introducing coefficients $P_{l}, Q_{l}$ as
\begin{eqnarray}
P_{l} &=& -\frac{(E_{l}+\omega_{V})(E_{l}+\omega_{R}+\omega_{V})+
|g^{S}|^{2}-|g^{A}|^{2}}{2g^{S}\omega_{R}},\nonumber\\
Q_{l} &=& -\frac{g^{S}P_{l}+E_{l}+\omega_{V}}{
g^{A\ast}}, \nonumber
\end{eqnarray}
we write the field operators as
\begin{eqnarray}
\hat{a}_{S}^{\dagger}(t) &=& \sum_{l} P_{l} A_{l} e^{iE_{l}t},\\
\hat{a}_{A}(t) &=& \sum_{l} Q_{l} A_{l} e^{iE_{l}t}. \nonumber
\end{eqnarray}
Common operator coefficients $A_{l}$ are determined in
terms of the operators $a_{V}(0),a^{\dagger}_{S}(0),
a_{A}(0)$  using the Cramer's rule
$\hat{{\cal A}}_{l} = \det{(D_{l})}/\,\det{(D)}$ where
\begin{eqnarray}
D= \left( \begin{array}{ccc}
1     &   1     &   1      \\
P_{1} &  P_{2}  &  P_{3}   \\
Q_{1} &  Q_{2}  &  Q_{3}
\end{array} \right) .\nonumber
\end{eqnarray}
and $D_{l}$ is the matrix obtained by
replacing the elements in the $l$th column of $D$ by the column vector
$[\hat{a}_{V}(0),\hat{a}^{\dagger}_{S}(0),\hat{a}_{A}(0)]^{T}$.
Thus, parameters $u,v,w$ are determined in terms of interaction
constants and the frequencies. More explicit expressions are too long
and not very illuminating to reproduce here, but above analysis is
quite suitable for numerical computation when some experimental data
is available. At that moment we shall content ourselves with more
fundamental discussions only.

In order to give a brief discussion of the dependence of 
the correlation function in Eq.9 on squeezing parameter 
and temperature, we consider an equilibrium distribution of 
vibration mode as of the squeezed thermal state
with the following mean number and number variance\cite{kim}
\begin{eqnarray}
\langle n_{V} \rangle &=& \bar{n}_{V}\cosh{2r}+\sinh^{2}{r}, \nonumber\\
V_{0}(n_{V}) &=& (\bar{n}_{V}^{2}+\bar{n}_{V})\cosh{4r}+
\frac{1}{2}\sinh^{2}{2r},
\end{eqnarray}
where $\bar{n}_{V}$ is the mean number of phonons according to Bose-Einstein
(BE) distribution and $r$ is the real squeezing parameter. When $r=0$, we 
recover the usual BE-distribution.
According to Eq.10 the $S$ and $AS$ correlations increases with variance
of phonons. And since both the $n_{V}$ and the $V(n_{V})$ increases
with temperature due to Eq.12, we see that temperature enforces stronger
correlations of $S$ and $AS$ modes. However, we need to put
a word of caution here,
since the fluctuations which are determined by the self-correlations
of the modes also increases with the temperature.
In order to represent this competition, one can consider the
cross-correlation function defined by\cite{law}
\begin{eqnarray}
C_{S-AS}=\frac{\langle n_{S},n_{A}\rangle}{\sqrt{V(n_{S})V(n_{A})}}.
\end{eqnarray}
Since the denominator can be expressed in a similar structure as with the
correlation function in Eq.10, the cross correlation function will
eventually saturate at high temperatures and at high squeezing
parameters. Therefore, at high temperatures thermal fluctuations becomes
important but not more important than in any typical quantum measurement.
An estimation for a typical ionic crystal, for example,
shows that the level of quantum fluctuations of phonon number exceeds
that of thermal fluctuations below $30 \div 50$K \cite{shu98,shu91}.
We also see through Eq.10 and Eq.12 that $S$ and $AS$ correlation
increases with the squeezing parameter $r$.

Finally, we examine the time range of validity for the parametric
approximation. For that aim,
we consider the Hamiltonian given in (1) for the case of perfect
coupling of single modes.  Let us suppress the momentum within the
mode labels $R,S,V,A$ and  calculate $a_{R}(t)$ for times close to
the beginning of interaction\cite{per2}.  Up to the second order, we
get
\begin{eqnarray}
a_{R}(t)=e^{-i\omega_{R}t}(a_{R}+it(M^{S*}a_{S}a_{V}+M^{A*}a_{A}
a_{V}^{\dagger})-\frac{1}{2}t^{2}(|M^{S}|^{2}\nu+|M^{A}|^{2}\mu))
\end{eqnarray}
where $\nu=a_{R}(n_{S}+n_{V}+1), \mu=a_{R}(n_{A}-n_{V})$. Here, operators
at $t=0$ are those without time arguments. Then, we calculate the mean
number and the variance of pump photons for S and AS modes are in vacuum
states initially as
\begin{eqnarray}
n_{R}(t)&=&n_{R}-t^{2}(|M^{S}|^{2}n_{R}(1+n_{V})+|M^{A}|^{2}n_{R}n_{S}),
\\\nonumber
V(n_{R}(t))&=&V(n_{R})+2t^{2}(|M^{S}|^{2}(V(n_{R})(1+n_{V})+
n_{R}(1+n_{V})+
\\\nonumber
&+&|M^{A}|^{2}(V(n_{R})n_{V}-n_{R}n_{V})))
\end{eqnarray}
In these equations averaging symbol, $<.>$, is not shown. Using the
relation \mbox{$V(n_{R})=n_{R}$} for a coherent field, we find the time 
ranges $t << \tau_{1}, \tau_{2}$, for which the field intensity and the 
variance remain close to their initial values, as
\begin{eqnarray}
\tau_{1} &=& \frac{1}{|M^{S}|^{2}(1+n_{V})+|M^{A}|^{2}n_{V}},\\\nonumber
\tau_{2} &=& \frac{1}{4|M^{S}|^{2}(1+n_{V})}.
\end{eqnarray}
Clearly, we see a rescaling of time range of the usual time range of
parametric approximation. At low temperatures $n_{V} \approx 0$ and
thus $\tau_{2}=(1/4) \tau_{1}$ shows a reduction of time range to $1/4$
of the typical range of parametric approximation. As an estimation, we may
take $g^{S} \approx 10^{7}Hz$\cite{per2}, giving time ranges as
$\tau_{1}=10fs$ and $\tau_{2}=2.5fs$. These ranges are readily available 
due to the remarkable recent developments in the field of femto-second
spectroscopy \cite{die,sha}.

\section{conclusion}

Summing up our results, we should stress that the measurement of
Stokes--anti-Stokes correlations looks like a reasonable method for
detecting the number variance of a Raman-active vibration mode
in solids. The most interesting and crucial fact is that the above 
method permits us to determine the number variance at thermal 
equilibrium, in other words, the variance just before the 
application of the pump beam.
The phonon sub-system could be be in a non-classical state due to an  
interaction providing necessary correlations among phonons before the
pump beam is applied.
That interaction could be some
anharmonic coupling with the heat bath, polaron or polariton
mechanisms. Since these mechanisms are usually weaker than the 
first order Raman effect, 
after the application of the pump beam, dynamics of the phonon
system is governed mainly by the Raman effect. Therefore, initial
non-classical state of phonons  and non-classical effects
like squeezing which require phase coherence might be destroyed.
That is why we have determined the general and
fundamental formula given by Eq.4 in terms of the initial state
of phonons and showed that under certain conditions it provides
direct information on the initial, thermal equilibrium variance
of phonons.
Analyzing those conditions of applicability,
we propose that
at liquid $N$ temperatures, using an intense coherent beam of
ultra-fast laser source such as $Ti$-sapphire as a pump for a Raman
active medium, one can measure the number correlation of the
scattered Stokes and anti-Stokes modes and the mean photon numbers
in these modes simultaneously by some photon counters, 
in order to determine the 
number variance of the vibration mode at equilibrium.
The measurement can be realized through the use of a homodyne-type
scheme\cite{man} in which the $S$ and $AS$ photons are counted by
two different detectors connected with a computer fixing the 
simultaneous arrival of the $S$ and $AS$ photons.
It is also shown that when the vibration mode is in squeezed
state then an increase in the correlation of the Stokes and       
anti-Stokes modes occurs. 

Case of a multi-mode pump, important for ultra-short pulses,
can be handled easily for materials which involves a
strongly preferred phonon mode due to a Van Hove singularity
in the frequency range of the pump, by an appropriate 
calculation of the effective coupling constants defined
by Eq.7 which in turn modify only the coefficients 
$A^{\prime},B^{\prime},C^{\prime}$ in Eq.12. Thus
our conclusions should also be valid in this case. 
For materials in which such phonon modes are many or 
not exists at all, then application of a multi-mode pump and 
measurement of Stokes-Anti-Stokes correlation would still 
provide information on multi-mode phonon correlations according 
to the general formula Eq.11.
This is a valuable knowledge to classify a possible non-classical
multi-mode state of phonons like a multi-mode squeezed state.

So far, the best achievement in squeezing of phonons is reported
to be $0.01\%$ \cite{gar}, provided by second order Raman scattering.
We would like to emphasize that this is not the squeezing
parameter $r$ of Eq.12 but related to $V(n_V)$. Hence, 
the change in the Stokes--Anti-Stokes correlations we expect 
to be in the same order. There are other mechanisms which
result in non-classical excitations in solids with different
expressions and larger values for $r$ and $V(n)$. In fact, 
squeezing parameter reflects the strength of interaction preparing 
the non-classical state of these excitations\cite{hu}, 
which is the initial phonon state in our scheme.
The example of optical polariton we have discussed in the introduction,
provides a two-mode squeezed state with squeezing parameter in the 
range $r \sim 0.1 - 0.01$ in $CuCl$\cite{art}. 
Therefore, such a measurement with the ultrafast 
Raman correlation spectroscopy should not be too challenging 
and looks  promising in our opinion. 

Let us finally note that the case of molecular Raman spectroscopy 
can also be treated with a similar formalism to get information
on the quantum statistics of populations of molecular energy levels.
 
\section{ACKNOWLEDGMENTS}
We acknowledge the useful discussions with Prof. A. Bandilla, and
Prof. V. Rupasov.

\begin{figure}
\caption{Phonon degree of coherence $G^{(2)}$ versus temperature for typical 
parameters of an ionic crystal: $\Omega =200K$, $g=25K$.}
\end{figure}

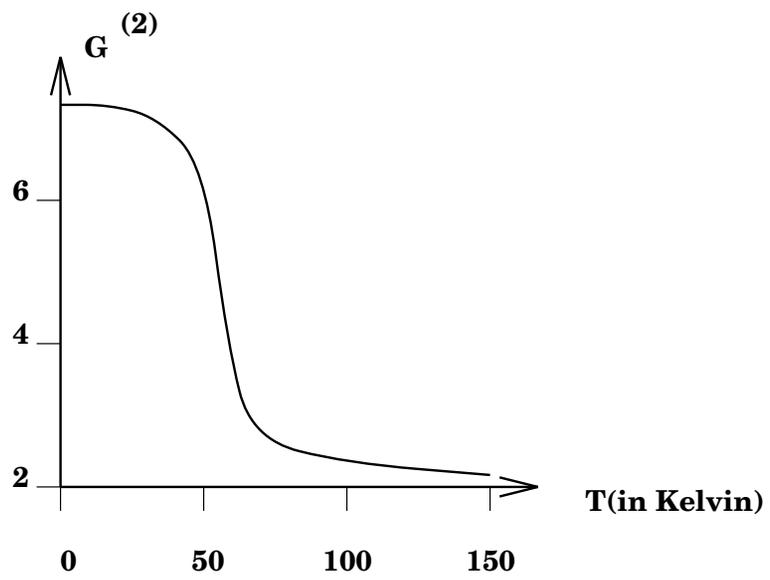
\end{document}